\documentstyle[preprint,aps]{revtex} 
\begin{document} 
\draft

\title{Do static sources outside a Schwarzschild black hole radiate?}


\author{Atsushi Higuchi}
\address{Institut f\"ur theoretische Physik, Universit\"at Bern,\\
          Sidlerstrasse 5, CH--3012 Bern, Switzerland\\ and\\
         Department of Mathematics, University of York,\\ 
         Heslington, York YO1 5DD, United Kingdom}
\author{George E.A. Matsas}
\address{Instituto de F\'\i sica Te\'orica, 
         Universidade Estadual Paulista,\\
         Rua Pamplona 145,
         01405-900-S\~ao Paulo, S\~ao Paulo,
         Brazil}
\author{Daniel Sudarsky}
\address{Instituto de Ciencias Nucleares,
         Universidad Nacional Aut\'onoma de M\'exico,\\
        A. Postal 70/543, M\'exico D.F. 04510,
       Mexico}
\date{\today}
\maketitle 


\begin{abstract}
We show that static sources coupled to a massless scalar field in 
Schwarzschild spacetime give rise to emission and absorption of zero-energy
particles due to the presence of Hawking radiation.  This is in complete
analogy with the description of the bremsstrahlung by a uniformly accelerated
charge from the coaccelerated observers' point of view.  The response rate
of the source is found to coincide with that in Minkowski spacetime as a
function of its proper acceleration.  This result may be viewed as restoration
of the equivalence principle by the Hawking effect.
\end{abstract} 

\narrowtext 


The relation between radiation from accelerated charges 
and the equivalence 
principle has for some time been the 
source of much confusion and discussion. 
A particularly interesting question is how to reconcile 
the following two facts (in Minkowski spacetime): On the one hand, 
an accelerated charge 
is known to radiate 
when it is seen from the viewpoint of inertial
observers.  On the other hand,
according to the equivalence principle, the
 same charge is 
seen by comoving observers as a static charge in
a uniform ``gravitational field", and, hence, is not expected to radiate.
In the classical context, this question has been answered first
by Rohrlich\cite{Roh} and further clarified by
Boulware\cite{B}, who has shown 
that the presence of a horizon for the 
collection of
comoving observers, who perceive the charge as static, 
serves to  explain 
the apparent paradox. 
This resolution is based on  the fact that the radiation zone (as described by
the Minkowski observers)
lies beyond the comoving observers' horizon and is 
thus unobservable by them.
In the  quantum 
mechanical context, a solution to the apparent paradox 
(which is now  cast in terms  of photon emission
rates) has been 
given by the authors\cite{HMS}, by recalling that, as
seen by the comoving observers,
the static charge (which has in fact constant proper 
acceleration)
is immersed  in the Fulling-Davies-Unruh (FDU) thermal bath 
\cite{D,U,F} in Rindler spacetime\cite{Rind}.
That is, the interaction of the static charge
with this thermal bath results 
in the absorption and stimulated emission 
of photons with zero Rindler energy
and this completely accounts for 
the bremsstrahlung due to a uniformly accelerated charge in quantum
electrodynamics.
(Here, the Rindler energy means
the energy corresponding to the boost Killing vector field with respect to
which Rindler spacetime is static.)

The purpose of this Letter is to note that, in complete
analogy to the result
obtained in the case of the static charge in Rindler
spacetime as described before,
the analysis of  a static
charge in a static black-hole spacetime, which interacts with 
Hawking radiation\cite{H}, yields 
a finite response rate.
In fact we will see that the total response
rate is {\em exactly} the same as that of a uniformly accelerated
source in Minkowski spacetime as a function of the proper acceleration.

We first review the general formalism for  
computing the response rate of 
a classical source in a static spacetime and
the result of Ref.\cite{HMS} in the context of massless scalar
field\cite{RW}.  Then we present our result for Schwarzschild
spacetime.

Let us consider a globally-hyperbolic static spacetime 
described by the metric
$ds^2 = f({\bf x})dt^2 - h_{ij}({\bf x})dx^i dx^j$.
We will study a real scalar field $\Phi$ 
that interacts with a classical source $j(x)$ ($x = (t, {\bf x})$)
and is described by the action
$S = \int d^4 x \sqrt{fh}\,\left( {1 \over  2}
\nabla^{\mu}\Phi \nabla_{\mu} \Phi
 +j \Phi\right)$,
where $h({\bf x}) = {\rm det}\,h_{ij}({\bf x})$.
Let 
\begin{equation}
u_{\omega{\bf s}\lambda}(x)
= \sqrt{\frac{\omega}{\pi}}\,
U_{\omega {\bf s}\lambda}({\bf x})\exp(-i\omega t) \label{Udef}
\end{equation}
with $\omega > 0$ and their complex
conjugates,
$u_{\omega {\bf s}\lambda}(x)^{*}$, be solutions to 
$\Box u = 0$, 
where ${\bf s} = (s_1,\cdots, s_n)$ is a set of
continuous labels and $\lambda$ is a discrete label
for the complete set of modes. We have assumed $\omega$ to be 
continuous because this is the case in the spacetimes we study, 
and adopted it as one of the labels. The factor of
$\sqrt{\omega/\pi}$ has been inserted for later
convenience.  Let these solutions
be Klein-Gordon orthonormalized:
\begin{eqnarray}
& & i\int d\Sigma\, n^{\mu}\left( 
u_{\omega {\bf s}\lambda}^{*}
\nabla_{\mu}u_{\omega'{\bf s}'\lambda'}
- \nabla_{\mu}u_{\omega {\bf s}\lambda}^{*}\cdot 
u_{\omega' {\bf s}'\lambda'} \right) \nonumber \\
   & & = \delta(\omega - \omega')
\delta({\bf s}-{\bf s}')\delta_{\lambda\lambda'}, \label{KG} \\
 & & i\int d\Sigma\, n^{\mu}\left( 
u_{\omega{\bf s}\lambda}\nabla_{\mu}u_{\omega'{\bf s}'\lambda'}
- \nabla_{\mu}u_{\omega{\bf s}\lambda}\cdot 
u_{\omega'{\bf s}'\lambda'}\right)  =  0, 
\end{eqnarray}
where $d\Sigma$ is the volume element of a Cauchy surface and where
$n^{\mu}$ is the future-pointing unit normal to it. 
The in-field $\Phi^{\rm in}$ satisfying 
the free field equation $\Box \Phi^{\rm in} = 0$ can now be expanded as 
$$
\Phi^{\rm in}(x) = \sum_{\lambda} \int d\omega d^n{\bf s}
\left[ u_{\omega {\bf s}\lambda}(x) a^{\rm in}_{\omega {\bf s}\lambda} + H.c. 
\right].
$$
Let the initial state be the in-vacuum state 
$|0\rangle_{\rm in}$ defined by
$a^{\rm in}_{\omega{\bf s}\lambda}|0\rangle_{\rm in} = 0$ for all $\omega$,
${\bf s}$
and $\lambda$. 

We will be interested in static sources. However,  
as we will see later, we need to introduce oscillation as a
regulator in order to avoid the appearance of intermediate indefinite results.
Therefore we consider a source  of the form
$j_{\omega_0}(x) = J({\bf x})\cos\omega_0 t$. 
The rate 
of spontaneous emission with fixed ${\bf s}$ and $\lambda$ 
can now be found to lowest order in perturbation theory:
$$
R_{\rm sp}(\omega_0;{\bf s},\lambda)d^n{\bf s} = \frac{\omega_0}{2}
|\tilde{J}(\omega_0, {\bf s},\lambda)|^2 d^n{\bf s}, \label{Sprate}
$$
where $\tilde{J}(\omega_0,{\bf s},\lambda) = \int d^3{\bf x}\,
\sqrt{h({\bf x})f({\bf x})}\,
J({\bf x})
U_{\omega_0 {\bf s}\lambda}({\bf x})$.
We note that Eq.\ (\ref{Sprate}) gives the emission
rate per unit {\em coordinate} time. Later we will convert it
into the rate per unit {\em proper} time for point sources. 

If the source is immersed
in a thermal bath of inverse temperature
$\beta = 1/k_B T$, the rates of absorption and {\em induced} emission
are both $R_{\rm sp}(\omega_0;{\bf s},\lambda)/(\exp\beta\omega_0 - 1)$.
Summing 
the absorption rate and the spontaneous and induced emission rates,
we find the total {\em response} rate:
$$
R(\omega_0;{\bf s},\lambda)
= \frac{\omega_0}{2}\coth\frac{\beta\omega_0}{2}
|\tilde{J}(\omega_0,{\bf s},\lambda)|^2 .
$$
In the case of interest here, i.e. for $\omega_0 \to 0$, we have 
\begin{equation}
R(0;{\bf s},\lambda)
= \beta^{-1}|\tilde{J}(0,{\bf s},\lambda)|^2. \label{Maineq}
\end{equation}

Let us now 
review how the bremsstrahlung rate due to a uniformly 
accelerated source (in Minkowski spacetime) is 
reproduced from the Rindler-spacetime point of view
by  taking the FDU thermal bath into account. 
First we present the
conventional result for the emission rate, which is to be compared with
the  Rindler-spacetime result.
We define the Rindler coordinates $\tau$ and $\xi$
in terms of the usual
Minkowski coordinates  by 
$t  = a^{-1}e^{a\xi}\sinh a\tau$,
$z  = a^{-1}e^{a\xi}\cosh a\tau$,
and consider the classical source
$j_0 = q\delta(\xi)\delta(x)\delta(y)$.
This source has constant
proper acceleration $a$.
Using the standard method (see, e.g., Ref.\cite{BD}), 
we obtain
the rate of spontaneous emission of particles with fixed transverse momentum
$(k_x,k_y)$ :
\begin{eqnarray}
R_{\rm sp}^M(k_x,k_y)dk_x dk_y 
 &  = & \int_{-\infty}^{+\infty}
dw\Delta_{k_{\perp}}\left(\frac{2}{a}\sinh\frac{aw}{2}\right)
\frac{dk_x dk_y}{(2\pi)^2} \nonumber \\
 & =  & \frac{1}{4\pi^3 a}[K_0 (k_{\perp}/a)]^2 dk_x dk_y, \label{Resmink}
\end{eqnarray}
where $k_{\perp} = \sqrt{k_x^2 + k_y^2}$. (We refer the reader to
Ref.\cite{GR} 
for formulas involving special functions used in this Letter.) 
The function
$\Delta_{m}(\sqrt{\sigma}) = -\frac{1}{4}N_0(m\sqrt{\sigma})$, 
where $\sigma = t^2 - z^2$,
is the symmetrized two-point function of massive scalar field
in two dimensions with $\sigma > 0$.

We can now compare the rate (\ref{Resmink})  with the
rate obtained in the Rindler point of view, where the 
variable $\tau$ is adopted as time. We first note
that  from this perspective the source is immersed 
in the FDU thermal bath.
This source absorbs particles from the heat bath, which
also gives rise to
induced emission.  Since the particle concept depends on the timelike 
Killing vector
that one uses to define it, emission of a Minkowski particle (i.e. one defined
with respect to $\partial/\partial t$)
can correspond either to absorption or to emission of a Rindler particle 
(i.e. one defined with respect
to $\partial/\partial \tau$) \cite{UW}.
However, the rate of
{\em response}, i.e. {\it emission plus absorption},
must be independent of the description that one uses.  
Therefore, the rate of spontaneous emission
given by (\ref{Resmink}) should equal the total response rate of the source
$j_0$ computed in Rindler spacetime with the 
FDU thermal bath.

 There is a technical complication with the verification 
of the above statement due to 
the fact that 
the spontaneous emission rate vanishes because the
source is now static whereas the density of states in the
thermal bath diverges in the zero-frequency limit.
As a result, we encounter
an expression of the form
$0\times \infty$ in the process of computing the 
response rate using the particle concept in 
Rindler spacetime.  For this reason we  regularize the 
calculation by
considering
\begin{equation}
j = \sqrt{2}\,q\cos\omega_0 \tau\,
\delta(\xi)\delta(x)\delta(y) \label{Rsource}
\end{equation} 
and taking the limit $\omega_0 \to 0$ in the end.  The factor of $\sqrt{2}$ is
necessary to make 
the time average of the squared charge equal $q^2$.
The source 
(\ref{Rsource}) is then equivalent  
to the source $j_0$ 
in the limit $\omega_0\to 0$ 
because the rate is proportional
to the squared charge at the lowest order. 

Now we verify explicitly that the $\omega_0\to 0$ limit of the total response
rate of the source (\ref{Rsource}), which is obtained from (\ref{Maineq})
with $\beta^{-1} = a/2\pi$, coincides with the rate (\ref{Resmink}).
The positive-frequency
modes with respect to $i\partial/\partial\tau$ are given by
\begin{equation}
u_{\omega k_x k_y}(\tau,\xi,x,y) =\sqrt{\frac{\omega}{\pi}}\,
\psi_{\omega k_{\perp}}(\xi) \times 
\frac{e^{ik_x x + ik_y y - i\omega \tau}}{2\pi}, \label{Uinpsi}
\end{equation}
where
\begin{equation}
\left[ -\frac{d^2\ }{d\xi^2} + k_{\perp}^{2}e^{2a\xi}\right]
\psi_{\omega k_{\perp}}(\xi) = \omega^2\psi_{\omega k_{\perp}}(\xi),
\label{MObessel}
\end{equation}
and where $k_{\perp} = \sqrt{k_x^2 + k_y^2}$. 
Requiring that $\psi_{\omega k_{\perp}}(\xi)$ 
decrease for $\xi \to +\infty$, we find that
$\psi_{\omega k_{\perp}}(\xi) \propto K_{i\omega/a}((k_{\perp}/a)e^{a\xi})$.
By the usual method of turning the
normalization integral into a surface term (see, e.g., Ref.\cite{HMS}), 
we find that the function
$u_{\omega k_x k_y}$ is normalized according to (\ref{KG}) if for large
and negative $\xi$
\begin{equation}
\psi_{\omega k_{\perp}}(\xi) \approx -\frac{1}{\omega}\sin 
[\omega \xi + \alpha(\omega)]. \label{Asymptotic}
\end{equation}
This determines $\psi_{\omega k_{\perp}}(\xi)$ :
$$
\psi_{\omega k_{\perp}}(\xi) 
= \sqrt{\frac{\sinh(\pi\omega/a)}{\pi a \omega}}
K_{i\omega/a}((k_{\perp}/a)e^{a\xi}).
$$
Consequently, we find
\begin{equation}
\psi_{0 k_{\perp}}(\xi) = a^{-1}
K_{0}((k_{\perp}/a)e^{a\xi}). \label{Zero}
\end{equation}

We note here that 
$\psi_{0 k_{\perp}}(\xi) \approx - \xi + {\rm const}.$ for large and
negative $\xi$.
This can be understood as the $\omega \to 0$ limit  of (\ref{Asymptotic}).
In fact, one can directly determine
the normalization factor of $\psi_{0 k_{\perp}}$ by requiring
this behavior
without referring to the solutions with nonzero $\omega$.  We will
use this method for the 
Schwarzschild black-hole case. 

Using (\ref{Uinpsi}) with
(\ref{Zero}) in (\ref{Maineq}), one finds that the total {\em response}
rate in the thermal bath of temperature $\beta^{-1} = a/2\pi$ in Rindler
spacetime is indeed equal to $R_{\rm sp}^M(k_x,k_y)$ given by
(\ref{Resmink}).
We compute the integrated response rate
 given by the integral over the transverse momentum for later use:
\begin{equation}
R_{\rm sp}^{M,{\rm tot}} = \int dk_x dk_y R_{\rm sp}^M(k_x, k_y)
= \frac{q^2}{4\pi^2} a. \label{Flattotal}
\end{equation}
 
Now we turn our attention to the Schwarzschild case.
In particular, we determine the response rate
of a point source analogous to (\ref{Rsource}) in the limit
$\omega_0 \to 0$. 
We use the standard Schwarzschild metric,
$ds^2 = f(r)
dt^2 - f(r)^{-1}dr^2 - 
r^2 ( d\theta^2  +\sin^2 \theta d\varphi^2)$,              
where
$f(r) = 1-2M/r$.
The positive-frequency solutions 
to the massless scalar field equation in 
this spacetime can be written as 
\begin{equation}
u_{\omega lm} = \sqrt{\frac{\omega}{\pi}}\, 
\frac{\psi_{\omega l}(r)}{r}
\times Y_{lm}(\theta,\varphi)
e^{-i\omega t}.
\label{SKG}
\end{equation}
Here $\psi_{\omega l}(r)$ is the 
solution to the differential equation
\begin{equation}
 \left\{ -f(r)
\frac{d}{dr}\left[f(r)
\frac{d\ }{dr} \right] + V_{\it eff}(r)\right\} 
\psi_{\omega l}(r)
= \omega^2 \psi_{\omega l}(r),
\label{RPWE}
\end{equation}
where
$V_{\it eff}(r) = \left( 1-2M/r\right)
\left[ 2M/r^3 + l(l+1)/r^2\right]$.
For given $\omega$, $l$ and $m$ there are two 
independent and orthogonal solutions of (\ref{RPWE}). One  
is purely incoming from the past horizon $H^-$ 
and the other is 
purely incoming from past null infinity ${\cal J^-}$.

In the Unruh vacuum\cite{U},
which corresponds to the physical black hole formed by
gravitational collapse, a thermal flux of temperature $\beta^{-1} = 1/8\pi M$
comes out from $H^{-}$.  In the Hartle-Hawking vacuum\cite{HH}
there is an additional
thermal flux coming from ${\cal J}^{-}$.  We concentrate on
the Unruh vacuum in this Letter. 

The regularized classical source we consider is 
\begin{equation}
j(x)  =  \frac{\sqrt{2}qf(r_0)^{1/2}}{r_0^2\sin\theta_0}
\cos\omega_0 t\, \delta(r-r_0)\delta(\theta-\theta_0)
\delta(\varphi - \varphi_0).
\label{OC}
\end{equation}
This source and the source (\ref{Rsource}) have the same strength
in the sense that they give the same value when integrated over the 
hypersurface of constant time.

Using Eq.\ (\ref{Maineq}) and 
 introducing
the correction factor $f(r_0)^{-1/2}$ to convert the rate per 
{\em coordinate} time into that per {\em proper} time, 
we find that the response rate per
{\em proper} time of the source (\ref{OC}) with fixed angular momentum
in the 
limit $\omega_0\to 0$ is given by
\begin{equation}
R_{lm} = \frac{q^2}{4\pi M r_0^2}f(r_0)^{1/2}
|\psi_{0l}(r_0)|^2 |Y_{lm}(\theta_0,\varphi_0)|^2, \label{Rate}
\end{equation}
provided that the function $u_{\omega l m}$ in (\ref{SKG}) is normalized
according to 
(\ref{KG}). 
Now our task is to find the function $\psi_{0l}$ incoming from $H^-$
and corresponding to this normalization.
(Strictly speaking, we need to prove that 
$\psi_{\omega l}(r) \to \psi_{0l}(r)$ as $\omega \to 0$.)

It is useful to introduce the
dimensionless Wheeler tortoise coordinate
$x = y +\ln (y-1)$, where 
$y = r/2M$.  Eq.\ (\ref{RPWE}) can then be rewritten as
\begin{equation}
\left[ -{{d^2} \over {d x^2}} + 
 (2M)^2 V_{\it eff}(x)\right]\psi_{\omega l} 
= (2M\omega)^2\psi_{\omega l}.
\label{KG2}
\end{equation}
In the limit $\omega \to 0$, the incoming wave from the
white-hole horizon
is totally reflected towards the black-hole horizon.
This implies that the Klein-Gordon normalization (\ref{KG})
is achieved for 
$u_{\omega lm}$ with $M\omega \ll 1$ if 
$\psi_{\omega l}\approx 
-\omega^{-1}\sin[2M\omega x+ \alpha(\omega)]$ for large and negative
$x$.
Thus, in the limit $\omega \to 0$, we must normalize the solution
$\psi_{0l}$ so that
\begin{equation}
\psi_{0l} \approx - 2M x + {\rm const}.\ \ \ \ (x < 0,\ |x| \gg 1).
\label{Norma}
\end{equation}

Now Eq.\ (\ref{RPWE}), or equivalently Eq.\ (\ref{KG2}),
can be solved explicitly for $\omega = 0$.
The general solution is
$\psi_{0l}(y) = C_1 yP_l(2y-1) + C_2 yQ_l(2y-1)$, where
$P_l(z)$ and $Q_l(z)$ are Legendre functions 
of the first and second kinds
with the branch cut $(-\infty, 1]$ for $Q_l(z)$.
Note that $P_l(z)\sim z^l$ and $Q_l(z)\sim z^{-l-1}$ for large $z$ and 
that the solution we seek  must decrease for large $y$
since the wave is totally reflected back to the horizon. From these facts 
and the condition (\ref{Norma}) we 
find $\psi_{0l} = 4MyQ_l(2y-1)$. 
Substituting this in (\ref{Rate}), we have 
$$
R_{lm} = \frac{q^2}{\pi M}f(r_0)^{1/2}
[Q_l(z_0)]^2 |Y_{lm}(\theta_0,\varphi_0)|^2,
$$
where $z_0 = r_0/M-1$. It is possible to sum over $l$ and $m$ using
the formulas
$\sum_{m=-l}^{l}|Y_{lm}(\theta,\varphi)|^2 = (2l+1)/4\pi$ and 
$\sum_{l=0}^{\infty}(2l+1)[Q_l(z)]^2 = 1/(z^2 -1)$.
The result is 
\begin{equation}
R_{\rm tot} = \sum_{l,m}R_{lm}
= \frac{q^2}{4\pi^2}a(r_0), \label{Response}
\end{equation}
where $a(r_0) = Mf(r_0)^{-1/2}/r_0^2$ is the proper acceleration of the 
static source. Note that this is identical with (\ref{Flattotal}) as a
function of proper acceleration.

We have not rigorously proved the validity of our approach where 
we directly work with
the $\omega = 0$ modes satisfying the normalization condition (\ref{Norma}) 
instead of explicitly
taking the $\omega \to 0$ limit of the modes with $\omega \neq 0$.  However,
the exact agreement of (\ref{Flattotal}) and (\ref{Response}) itself 
and the fact that we
have reproduced with this method precisely the results of
Ref.\cite{RW} serve as consistency checks of our approach.
We will present elsewhere\cite{HMS2} a detailed analysis about how
the functions 
$\psi_{\omega l}$ approach $\psi_{0l}$ 
in the $\omega \to 0$ limit.  
Here we will present another consistency check using 
a model\cite{HMS3} where
the effective potential $V_{\it eff}$ is replaced by
a simpler but similar potential
$V^{\it (s)}_{\it eff}(x) = l(l+1)\theta(x-1)/(2Mx)^2$ ($l \neq 0$).   
With this replacement, the function $\psi_{\omega l}^{\it (s)}$
incoming from $H^-$
and corresponding to $\psi_{\omega l}$ can be found 
explicitly for any value of $\omega$, and we have 
\begin{eqnarray}
\psi^{\it (s)}_{\omega l}(x) 
 & = & a_{\omega l} (\beta^{(+)}_{\omega l}e^{i\tilde{\omega} x}
+ \beta^{(-)}_{\omega l}e^{-i\tilde{\omega} x})\ \ \ (x < 1), \nonumber \\
& = & a_{\omega l} xh_{l}^{(1)}(\tilde{\omega} x)\ \ \ \ \ \ \ \ \ \  
\ \ \ \ \ \ \ \ \ \ (x > 1), \nonumber
\end{eqnarray}
where $\tilde{\omega} = 2M\omega$.
Continuity of the value and the first derivative at
$x = 1$ gives
$\beta^{(\pm)}_{\omega l}  =  (1/2)e^{\mp i\tilde{\omega}}
[(1\mp i/\tilde{\omega})h_{l}^{(1)}(\tilde{\omega})
\mp i(h_l^{(1)})'(\tilde{\omega})]$.
The normalization condition leads to
$a_{\omega l}\beta^{(+)}_{\omega l} = i/2\omega$
up to a phase factor.  Then the mode functions with
$M\omega \ll 1$ can be approximated by 
\begin{eqnarray}
\psi^{\it (s)}_{\omega l}(x)
& = & 2M(1-x + l^{-1}) + O(\omega^2) \ \ \ (x < 1),
\nonumber \\
   & = & 2Ml^{-1}x^{-l} + O(\omega^2) \ \ \ \ \ \ \ \ \ \ \ \ (x > 1).
\nonumber
\end{eqnarray}
Thus, we see explicitly that the $\omega \to 0$ limit of
$\psi_{\omega l}^{\it (s)}(x)$ is indeed $\psi_{0l}^{\it (s)}(x)$ 
satisfying the
normalization condition (\ref{Norma}).

Note that the response rate of a static source will vanish in the absence of
the Hawking effect, i.e. in the Boulware vacuum\cite{Boul}, whereas the source
in Minkowski spacetime with the corresponding acceleration radiates.
In this sense the equivalence principle is violated in the Boulware vacuum.
Since the Hawking effect is closely related to the absence of singularity 
of the quantum state
on the future horizon\cite{FH,KW}, it is reasonable to 
expect that this effect
plays a crucial role in ``restoring the 
equivalence principle" near the horizon.
It is therefore not surprising that the rate of response (\ref{Response}) 
agrees with the corresponding result
(\ref{Flattotal}) in Minkowski spacetime in the limit $r_0\to 2M$. 
However, the fact that they coincide for all $r_0$ was rather unexpected.
It would be
interesting to see if this fact, i.e. the complete ``restoration of the
equivalence principle" for a static scalar source by the Hawking effect,
is a special case of a more general
phenomenon.  


\acknowledgments

We thank Bob Wald and Bernard Kay for useful discussions.
We also thank Chris Fewster 
for helpful comments 
on the zero-energy limit of one-dimensional scattering theory
and Mike Ryan for useful comments on the manuscript.
The work of AH was supported in part by 
Schweizerischer Nationalfonds and the Tomalla Foundation.
GM would like to acknowledge partial support from
the Conselho Nacional de Desenvolvimento Cient\'\i fico e
Tecnol\'ogico.
DS would like to acknowledge partial support from 
DGAPA-UNAM Project No. IN 105496.


\end{document}